\documentclass[12pt]{article}
\usepackage{url,graphicx,amsmath,amssymb,latexsym,subfigure}
\newtheorem{theorem}{Theorem}
\newtheorem{lemma}{Lemma}
\newtheorem{remark}{Remark}

\newtheorem{proposition}{Proposition}

\begin{document}
%\pagestyle{fancy}
%\renewcommand{\baselinestretch}{1.6}
%\lhead[\fancyplain{} \leftmark]{}
%\chead[]{}
%\rhead[]{\fancyplain{}\rightmark}
%\cfoot{}
%\headrulewidth=0pt

\vspace{0.8pc}
\centerline{\large\bf Functional Partial Linear Model}
\vspace{.4cm}
\centerline{Heng Lian}
\vspace{.4cm}
\centerline{\it Division of Mathematical Sciences,}
\centerline{\it School of Physical and Mathematical Sciences,}
\centerline{\it Nanyang Technological University,}
\centerline{\it Singapore 637371.}
\vspace{.55cm}

\begin{quotation}
\noindent {\it Abstract:}
When predicting scalar responses in the situation where the explanatory variables are functions, it is sometimes the case that some functional variables are related to responses linearly while  other variables have more complicated relationships with the responses. In this paper, we propose a new semi-parametric model to take advantage of both parametric and nonparametric functional modeling. Asymptotic properties of the proposed estimators are established and finite sample behavior is investigated through a small simulation experiment.\par

\vspace{9pt}
\par
\end{quotation}\par

\section{Introduction}
Since the introduction of the partial linear model by \cite{engle86}, it has been widely studied in the statistical literature \cite{hua93,fan99,you07,huang09,yang09,sun09}. Partial linear models belong to the class of semi-parametric models since they contain both parametric and nonparametric components. On the one hand, it addresses the curse of dimensionality problem associated with completely nonparametric models and facilitates interpretation of the effect of the covariates associated with the linear part. On the other hand, they are more flexible than the standard linear regression when it is believed that some covariates are nonlinearly related to the independent variable.

On another direction of statistical research, there has recently been increased interest in the statistical modeling of functional data. In many experiments, functional data appear as the basic unit of observations. As a natural extension of the multivariate data analysis, functional data analysis provides valuable insights into these problems. Compared with the discrete multivariate analysis, functional
analysis takes into account the smoothness of the high dimensional covariates, and often
suggests new approaches to the problems that have not been discovered before. Even for
nonfunctional data, the functional approach can often offer new perspectives on the old
problem.

The literature contains an impressive range of functional analysis tools for various problems
including exploratory functional principal component analysis, canonical correlation
analysis, classification and regression. Two major approaches exist. The more traditional
approach, masterfully documented in the monograph \cite{ramsey05}, typically
starts by representing functional data by an expansion with respect to a certain basis, and
subsequent inferences are carried out on the coefficients. The most commonly utilized basis
include B-spline basis for nonperiodic data and Fourier basis for periodic data. Another line
of work by the French school \cite{ferraty02}, taking a nonparametric point of view, extends the traditional nonparametric techniques, most notably the kernel estimate, to the functional case. Some theoretical results are also obtained as a generalization of the convergence properties of the classical kernel estimate. Some recent advances in the area of functional regression include \cite{cai06,saidi08,wong08}.

In this paper, our aim is to combine the parametric and nonparametric approaches to functional regression resulting in functional partial linear models. We are aware of two other works that introduced partial linear regression in a functional context, the so-called semi-functional partial linear model \cite{perez06} and partial functional linear model \cite{shin09}. The former combines nonparametric functional model with a standard linear regression component, while the latter used a functional linear model together with a standard linear regression model. Both models have a functional component as well as a non-functional linear component. To the best of our knowledge, our work is the first study that combines the parametric and nonparametric approaches to functional regression in a functional semi-parametric model.  

  In the next section, we present our new model and construct estimators for both the parametric and nonparametric components based on principal component regression and Nadaraya-Watson kernel estimator. Then we derive some consistency and convergence rate results for the two components. In Section 3, we illustrate our methodology with a simulation study. Finally, in Section \ref{sec:conc}, we conclude our findings with a discussion. The technical proofs are collected in the Appendix.

\section{Funtional partial linear models}\label{sec:method}
In our functional partial linear regression model, the data triplets $\{X_i,T_i,Y_i\}_{i=1}^n$, which are independent and identically distributed (i.i.d.), are generated from the model
\begin{equation}\label{model}
Y_i=\int_0^1 b(s)X_i(s)\,ds+g(T_i)+\epsilon_i.
\end{equation}
Both $X_i$ and $T_i$ are random functions belonging to $H=L^2([0,1])$, the Hilbert space containing square integrable functions defined on the unit interval with inner product $\langle x,y\rangle=\int_0^1x(s)y(s)ds\;\forall x,y\in H$, $b\in H$ is the regression coefficient for the linear part and $g$ is a general continuous function on $H$ and the mean zero errors $\epsilon_i$ are independent of the functional covariates $\{X_i,T_i\}$. Note that for simplicity we assume $T_i$ and $X_i$ are both in $L^2$ while in fact we can assume that $T_i$ belongs to a more general vectorial topological space on which a semimetric is defined. See \cite{ferraty04,ferraty06} for more discussions on various possible semimetrics. We will use $\{X,T,Y\}$ to denote the generic random variables with distribution the same as $\{X_i,T_i,Y_i\}$ while the corresponding lower-case letters $\{x,t,y\}$ denote nonrandom values that the random variables can assume. To ensure identifiability, we do not put a scalar intercept term in the model since the intercept can be incorporated into the nonparametric component. We also assume $X$ is a mean zero process. 

To obtain estimators for both components, we get the following equation by computing the conditional expectation of (\ref{model}) on $T$:
\[E(Y|T)=\langle b, E(X|T)\rangle +g(T).\]
Subtracting the above equation from (\ref{model}) we get the model with only the linear component:
\begin{equation}\label{newmodel}
Y-E(Y|T)=\langle b, X-E(X|T)\rangle +\epsilon.
\end{equation}
As $E(Y|T_i)$ and $E(X|T_i)$ are unknown, we replace both expressions by Nadaraya-Watson kernel estimators with
\[E(Y|T_i)\approx \frac{\sum_jK(||T_i-T_j||/h)Y_j}{\sum_jK(||T_i-T_j||/h)},\]
\[E(X|T_i)\approx \frac{\sum_jK(||T_i-T_j||/h)X_j}{\sum_jK(||T_i-T_j||/h)},\]
where $K$ is the kernel function and $h$ is the bandwidth that typically converges to zero as $n$ goes to infinity. We use the notations $w_{ij}=K(||T_i-T_j||/h)/\sum_kK(||T_i-T_k||/h)$ and $w(t,T_i)=K(||t-T_i||/h)/\sum_jK(||t-T_j||/h)$ below for convenience.

With the kernel estimators plugged into (\ref{newmodel}), we have formally the following functional linear model 
\begin{equation}\label{newsamplemodel}
\tilde{Y_i}=\langle b, \tilde{X}_i\rangle+\epsilon_i,
\end{equation}
with $\tilde{Y_i}=Y_i-\sum_jw_{ij}Y_j$ and $\tilde{X_i}=X_i-\sum_jw_{ij}X_j$. Obviously (\ref{newsamplemodel}) is the sample version of (\ref{newmodel}). 

Following \cite{cardot99,hall07}, we define the second moment operator $S$ by
\[S=E[(X-E(X|T))\otimes(X-E(X|T))],\]
with the interpretation of $S$ to be a mapping from $H$ to $H$: $S(x)=E[\langle X-E(X|T), x\rangle(X-E(X|T))],\;\forall x\in H$.
We also define the cross second moment operator $\Delta$ by
\[\Delta=E[(X-E(X|T))(Y-E(Y|T))].\] The sample version of $S$ is $\hat{S}=n^{-1}\sum_i \tilde{X}_i\otimes\tilde{X}_i$ and $\hat{\Delta}$ can be defined similarly.

Using the Karhunen-Loeve expansion, we can write
\[S=\sum_{j=1}^\infty\lambda_j\phi_j\otimes\phi_j\]
and
\[\hat{S}=\sum_{j=1}^\infty\hat{\lambda}_j\hat{\phi}_j\otimes\hat{\phi}_j,\]
with $\lambda_1\ge \lambda_2\ge\cdots$ the eigenvalues and $\phi_1,\phi_2,\ldots$ orthonormal eigenvectors associated with $S$. Similarly for  $\hat{\lambda}_1\ge \hat{\lambda}_2\ge\cdots$ and $\hat{\phi}_1,\hat{\phi}_2,\ldots$ associated with the sample version operator $\hat{S}$.

From (\ref{newmodel}), we get $S(b)=\Delta$. If we expand different quantities in terms of the orthnormal system  $\{\phi_j\}$, we have the representations $b=\sum_jb_j\phi_j$, $\Delta=\sum_j\Delta_j\phi_j$, with relation $b_j=\Delta_j/\lambda_j$, which leads to the principal component analysis based estimator used in \cite{cardot99,cardot03,hall07}:
\[\hat{b}=\sum_{j=1}^m\hat{b}_j\hat{\phi}_j\]
where $\hat{b}_j=\langle\hat{\Delta},\hat{\phi}_j\rangle/\hat{\lambda}_j$ and $m\le n$ is the truncation level that trades off approximation error against variability, and $m$ typically diverges with $n$.

Finally, the nonparametric component $g$ can be estimated as
\[\hat{g}(t)=\sum_jw(t,T_j)(Y_j-\langle \hat{b}, X_j\rangle).\]

Next we study consistency and rate of convergence for the proposed estimators. Before doing that, we state a simple model identifiability result which only requires the positive definiteness of the operator $S$, which will be assumed throughout the paper.
\begin{proposition}\label{prop1}
Assume that the operator $S$ is positive definite (i.e. $\lambda_j>0\;\forall j$), then model (\ref{model}) is identifiable. Specifically, $E(Y|X,T)=\langle b_1,X\rangle+g_1(T)=\langle b_2,X\rangle+g_2(T)$ implies that $b_1=b_2$ and $g_1=g_2$ on the support of the distribution of $T$.
\end{proposition}

The assumptions required for our consistency result are stated as follows. 
\begin{enumerate}
\item[(A)] $\int E(X^4)<\infty, E\epsilon^3<\infty$ and the distribution of $T$ is supported on a compact subset of $L^2([0,1])$.
\item[(B)] $\lambda_1>\lambda_2>\lambda_3>\cdots>0$, i.e. the eigenvalues are positive with multiplicity $1$.
\item[(C)] The kernel $K$ satisfies the usual condition: $K$ has support $[0,1]$, continuous on $[0,\infty)$ and $-K'(u),u\in (0,1)$ is positive and bounded away from zero.  
\item[(D)] The function $g(t)$ in model (\ref{model}) and $h(t)=E(X|T=t)$ are Lipschitz continuous of order $\gamma$: $|g(t_1)-g(t_2)|\le C||t_1-t_2||^\gamma$, $||h(t_1)-h(t_2)||\le C||t_1-t_2||^\gamma$.
\item[(E)] The bandwidth $h$ satisfies $h\rightarrow 0$ and $n\phi(h)\rightarrow\infty$, where $\phi(h)$ is the asymptotic order of the so-called small ball probability, that is , $c_0\phi(h)\le P(||T-t||<h)\le c_1\phi(h)$ for some $c_0, c_1>0$ and for all $t$ in the support of the distribution of $T$.
\item[(F)] $m\rightarrow\infty, k_n^{-1}\lambda_m^2\rightarrow\infty$, where $k_n=h^\gamma+\sqrt{1/(n\phi(h))}$.
\item[(G)] $k_n^{-1}\lambda_m/(\sum_{j=1}^m\delta_j^{-1})\rightarrow\infty$ where $\delta_j=\min_{1\le k\le j}(\lambda_k-\lambda_{k+1})$.
\end{enumerate}
%Add discussions on assumption????

The consistency proof for the theorem below makes use of existing results for the functional linear model \cite{cardot99} but the assumptions we need are stronger due to the presence of the nonparametric component.
\begin{theorem}\label{consistency}
Suppose that assumptions (A)-(G) are satisfied, then $||\hat{b}-b||+|\hat{g}(t)-g(t)|\rightarrow 0$ in probability.
\end{theorem}
To calculate the rates of convergence, we make the following additional assumptions on the various Fourier coefficients defined previously:
\begin{enumerate}
\item[(H)]$\lambda_j-\lambda_{j+1}\ge C^{-1}j^{-\alpha-1}$, $|b_j|\le Cj^{-\beta}$ for some $C>0,\alpha>1,\beta>1$.
\end{enumerate}
\begin{theorem}\label{rate}
Under assumptions (A)-(H), we have the convergence rates (for convergence in probability) $||\hat{b}-b||^2=O_p(k_n^2m^{4\alpha+3}+m^{-2\beta+1})$, $|\hat{g}(t)-g(t)|^2=O_p(k_n^2m^{4\alpha+3}+m^{-2\beta+1})$.
\end{theorem}
\begin{remark}
With only the parametric component, \cite{hall07} showed that the optimal rate for $||\hat{b}-b||^2$ is $O_p(n^{-(2\beta-1)/(\alpha+2\beta)})$ if $\beta>1+\alpha/2$. With only the nonparametric component, \cite{ferraty06} obtained the rates $O_p(k_n^2)$ (if their results are adapted to the case of convergence in probability instead of almost surely). Our asymptotic results above show that in a functional partial linear model we can only obtain a substantially slower rate. Further discussions on this point are made in Section \ref{sec:conc}. 
\end{remark}

\section{Simulation}
In this section, we provide a numerical example to illustrate the methodology and theory presented previously. We simulate samples $(X_i,T_i,Y_i)$ from model (\ref{model}). For the linear component, we take $b=\sum_jb_j\phi_j$ with $b_1=0.5, b_j=4j^{-2}$ for $j\ge 2$, $\phi_1=1$, $\phi_j=\sqrt{2}\cos((j-1)\pi t)$ for $j\ge 2$ and $X=\sum_j\xi_ja_j\phi_j$ with $\xi_j$ independent and uniformly distributed on $[-\sqrt{3},\sqrt{3}]$ and $a_j=j^{-1}$. For the nonparametric component, we use 
\[g(t)=\int_0^1|t(s)|(1-\cos(\pi s))ds\]
and the random covariate curves for the nonparametric component are simulated marginally from 
\[T(s)=\sin(\omega s)+(a-\pi)s+d, \omega\sim Unif(0,2\pi), a,d\sim Unif(0,1).\]
To introduce some dependence between $X$ and $T$, we set $a=\xi_1/2\sqrt{3}+1/2$, $d=\xi_2/2\sqrt{3}+1/2$. Finally, Gaussian errors with standard deviations of $0.5$ are added to produce the final dependent variables.

To assess the performance of the procedure, we consider the following error criteria:
\begin{eqnarray*}
MSE_1&=&||\hat{b}-b||^2, \\
MSE_2&=&\sum_{i=1}^n(\hat{g}(T_i)-g(T_i))^2/n,\\
MSE_3&=&\sum_{i=1}^n(\langle \hat{b}-b,X_i\rangle+\hat{g}(T_i)-g(T_i))^2/n,
\end{eqnarray*}
which represent the errors for the functional linear coefficients, the nonlinear component of the regression function and the regression function respectively.

In the implementation, for the parametric linear component, we use B-spline of order $4$ with $20$ equi-spaced knots to represent the functional covariates with no additional smoothing (since no error is contained in the covariates). Functional principal component analysis is performed using the R package fda (\url{http://ego.psych.mcgill.ca/misc/fda/software.html}). For the nonparametric component, we use the quadratic kernel for the nonparametric estimator, with estimation performed using the npfda package (\url{http://www.math.univ-toulouse.fr/staph/npfda/index.html}). 

\begin{figure}
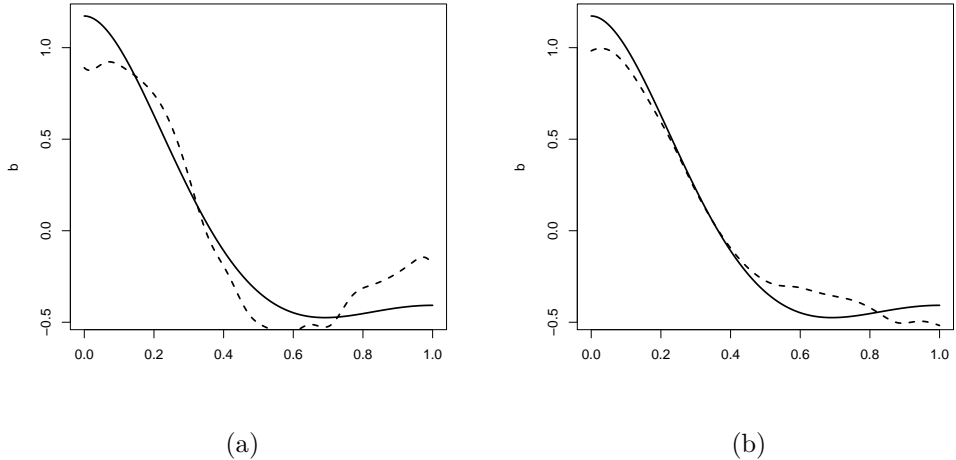

\centerline{\subfigure[]{\includegraphics[width=2.5in]{newn100.eps}
%\label{fig_second_case}
}
\hfil
\subfigure[]{\includegraphics[width=2.5in]{newn500.eps}
%\label{fig_second_case}
}}

\caption{Estimated functional linear coefficient (dotted line) with different sample sizes, (a)$n=100$; (b)$n=500$. \label{fig1}   }
\end{figure}

We present the simulation results for $n=100$ and $n=500$ in Table \ref{tab1} and \ref{tab2} respectively, with different truncation levels $m$ and different bandwidth parameters $h$. In the tables, the bandwidth $\hat{h}$ is the median of pairwise distances among the functional covariates, i.e., $\hat{h}=med_{i<j}\{||T_i-T_j||\}.$ For a given sample size $n$, our results represent averages over $100$ Monte Carlo replications for each parameter setting. The three numbers for each parameter setting correspond to the three error measures above. We note that for different error measures, the minimum errors are achieved at different parameter settings. We also show in Figure \ref{fig1} the estimated linear coefficient $\hat{b}$ using the optimal parameter settings (minimizing $MSE_1$) for both sample sizes. 

We then compare the performance of completely parametric and completely nonparametric estimators with the same data generated from the true model (\ref{model}). That is, we concatenate $X_i$ and $T_i$ and consider the new covariate as defined on the interval $[0,2]$ and then apply the two approaches for estimating the regression function. For these two estimators, only the mean squared error for the regression function ($MSE_3$) above makes sense, which is presented in Table \ref{tab3} and \ref{tab4} for the two estimators respectively. When the true model is partially linear, the completely linear model is clearly misspecified and results in extremely large mean squared errors. The completely nonparametric estimator is also not as good as the partial linear estimator since it loses some efficiency when $X$ is in fact linearly related to the responses. 
\begin{table}
  \caption{Simulation results (MSE) for our functional partial linear regression model when $n=100$. The minimum errors are emphasized with foldface font.\label{tab1}}
\bigskip
\centering{\begin{tabular}{ccccccc}\hline
   $m $ &$\hat{h}$ & $2\hat{h}$  & $4\hat{h} $ & $8\hat{h}$ &$16\hat{h} $     \\
\hline
   1 &0.068& 0.068 & 0.070&0.075&0.077\\
     &0.054& 0.050 &0.068 &0.181 &0.235\\
     &0.063& 0.062  &0.076 &0.202 &0.269\\
   2 & 0.014 & 0.014 & \textbf{0.012} &0.14&0.15 \\
     &0.047 & 0.044 & 0.072 & 0.104 &0.233\\
     &0.059 & 0.059 & 0.077 &0.142 & 0.275\\
   3&  0.018 & 0.020 & 0.029&0.035&0.040\\
     &0.044 & \textbf{0.040} & 0.068 & 0.131 &0.235\\
     &\textbf{0.051} & 0.053 & 0.080  & 0.153 &0.287\\
   4& 0.061  & 0.057 & 0.059 &0.063 & 0.074\\
     &0.045 & 0.042 & 0.060 & 0.133 &0.329\\
     &0.059 & 0.055 & 0.072 & 0.164 &0.391\\
   5& 0.106 &0.111 & 0.125 &0.215 & 0.227\\
     &0.044 & 0.063 & 0.079 & 0.141 &0.337\\
     &0.058 & 0.074 &0.100 & 0.178 & 0.411\\
%   6& 0.200 &0.202 & 0.154 & 0.264 &0.273\\
%     &0.048 & 0.063 & 0.076 & 0.138 &0.335\\
   \hline
  \end{tabular}}
      \label{symbols}
\end{table}

\begin{table}
  \caption{Simulation results (MSE) for our functional partial linear regression model when $n=500$.\label{tab2}}
\bigskip
\centering{\begin{tabular}{ccccccc}\hline
   bandwidth &$\hat{h}$ & $2\hat{h}$  & $4\hat{h} $ & $8\hat{h}$ &$16\hat{h}$      \\
\hline
   1 &0.0667& 0.0647& 0.0644&0.0645 &0.0646\\
     &0.0185& 0.0151&0.0447 &0.1367 &0.1622\\
     &0.0191& 0.0172&0.0481 &0.1525 & 0.1970\\
   2 &0.0045& 0.0039&0.0038 &0.0050 &0.0055 \\
     &0.0185& 0.0150&0.0449 &0.1287 &0.1454\\
     &0.0198& 0.0174&0.0490 & 0.1457& 0.1799\\
   3& 0.0064& \textbf{0.0035}& 0.0074&0.0062 &0.0061\\
     &0.0179& 0.0140&0.0428 &0.1093 &0.1367\\
     &0.0191& 0.0152&0.0473 & 0.1227& 0.1553\\
   4& 0.0073& 0.0045& 0.0093&0.0151 & 0.0165\\
     &0.0176& \textbf{0.0138}&0.0396 &0.1036 &0.1579\\
     &0.0191& \textbf{0.0147}&0.0415 & 0.1155& 0.1449\\
   5 &0.0178&0.0152 & 0.0146&0.0315 & 0.0376\\
     &0.0179& 0.0146&0.0405 &0.1154 &0.1834\\
     &0.0203& 0.0156&0.0499 & 0.1433& 0.2186\\
%   6& 0.0436 &0.0459 & 0.0467 & 0.0828 &0.0900\\
%     &0.040(min) & 0.044 & 0.168 & 0.304 &0.335\\
   \hline
  \end{tabular}}
      \label{symbols}
\end{table}

\begin{table}
  \caption{Simulation results (MSE) using data generated from the partial linear model but fitted using functional linear regression when $n=100$.\label{tab3}}
\bigskip
\centering{\begin{tabular}{cccccc}\hline
   $m $ &1 & 2  & 3 & 4 &5       \\
\hline
   &0.668& 0.573& 0.481& 0.617 &1.264\\
   \hline
  \end{tabular}}
      \label{symbols}
\end{table}

\begin{table}
  \caption{Simulation results (MSE) using data generated from the partial linear model but fitted using completely nonparametric regression  when $n=100$.\label{tab4}}
\bigskip
\centering{\begin{tabular}{cccccc}\hline
   bandwidth &$\hat{h}$ & $2\hat{h}$  & $4\hat{h} $ & $8\hat{h}$ &$16\hat{h} $     \\
\hline
   &0.106&0.089 & 0.138 &0.427& 0.578\\
   \hline
  \end{tabular}}
      \label{symbols}
\end{table}

\section{Conclusion}\label{sec:conc}
In this paper we initiate a study on functional partial linear models where both components are functional in nature. Consistency and convergence rates are obtained. Unlike the traditional partial linear model where the convergence rates for either component are the same under mild regularity conditions whether the other component is known or not, here for our functional model the rates obtained are worse than that of completely parametric or nonparametric models. From the proofs, this decrease in rate is caused by the convergence rate of $||\hat{S}-S||$ which in the completely parametric case is $O_p(1/\sqrt{n})$ \cite{cardot99,hall07}, while the unknown nonparametric component in our model makes the rate slower (Lemma \ref{lem1} in the Appendix). Although we do not have any corresponding lower bounds on the rates of convergence, it is reasonable to conjecture that the optimal rate cannot be achieved when the parametric component is infinite dimensional as in our functional model.

In our estimation procedure, we need to choose both the number of principal components for the parametric part and the bandwidth for the nonparametric part. Although we do not consider automatic selection for these parameters in the current study, we could use standard techniques such as K-fold cross-validation. With two parameters to search over, it is still to be seen whether we can get reasonable performances with limited computational resources.  Another open question is the construction of confidence bands for either the parametric or the nonparametric component. From a conceptual point of view, bootstrap method seems to be viable but its computational and theoretical properties remain as a challenge. All those problems deserve further investigations.

\section*{Appendix}
\textit{Proof of Proposition \ref{prop1}.}
If $E(Y|X,T)=\langle b_1,X\rangle+g_1(T)=\langle b_2,X\rangle+g_2(T)$, since $E(Y-\langle b_1,X\rangle-g_1(T))^2=E(Y-\langle b_2,X\rangle-g_2(T))^2+\langle S(b_1-b_2),b_1-b_2\rangle$, we have $b_1=b_2$ by the positive definiteness of $S$. Then $g_1=g_2$ follows from $g_j(T)=E[Y-\langle b_j,X\rangle|T], j=1,2$.

\bigskip
For any operator $U: H_1\rightarrow H_2$ which is a linear mapping between two Hilbert spaces, we consider the operator norm $||U||=\sup_{||x||_{H_1}\le 1}||U(x)||_{H_2}$. Note that there is no confusion when we use $||\cdot||$ for both the operator norm and the $L^2$ norm when $H_2$ is the real line because of the Riesz representation theorem. The following lemma gives the convergence rates for operators $\hat{S}$ and $\hat{\Delta}$.

\begin{lemma}\label{lem1}
Under the assumptions (A),(C)-(E) stated in Section \ref{sec:method}, we have
\[ ||\hat{S}-S||=O_p(k_n)\] and \[||\hat{\Delta}-\Delta||=O_p(k_n),\] 
where $k_n=h^\gamma+(n\phi(h))^{-1/2}$.
\end{lemma}
\textit{Proof.} By the definition of the operator $\hat{S}$, we have 
\begin{eqnarray*}
\hat{S}&=&\frac{1}{n}\sum_{i=1}^n(X_i-\sum_jw_{ij}X_j)\otimes(X_i-\sum_jw_{ij}X_j)\\
&=&\frac{1}{n}\sum_{i=1}^n(X_i-E(X|T_i))\otimes(X_i-E(X|T_i))\\
&&+\frac{1}{n}\sum_{i=1}^n(E(X|T_i)-\sum_jw_{ij}X_j)\otimes(X_i-E(X|T_i))\\
&&+\frac{1}{n}\sum_{i=1}^n(X_i-E(X|T_i))\otimes(E(X|T_i)-\sum_jw_{ij}X_j)\\
&&+\frac{1}{n}\sum_{i=1}^n(E(X|T_i)-\sum_jw_{ij}X_j)\otimes(E(X|T_i)-\sum_jw_{ij}X_j)\\
&=:&S_1+S_2+S_3+S_4.
\end{eqnarray*}
Lemma 5.2 in \cite{cardot99} showed that $||S_1-S||=O_p(n^{-1/2})=o_p(k_n)$. It can be shown that $\max_i||E(X|T_i)-\sum_jw_{ij}X_j||=O_p(k_n)$. The proof of this fact is similar to that of \cite{ferraty04,ferraty06} but is in fact simpler due to the fact that we only need to use Markov inequality to show convergence in probability instead of using Bernstein's inequality in showing almost sure convergence. The extra $\log n$ factor does not appear for the same reason when we are only interested in showing convergence in probability. Thus all three terms $S_2, S_3$ and $S_4$ are of order $O_p(k_n)$ and the rate for $||\hat{S}-S||$ is shown.  The proof for $||\hat{\Delta}-\Delta||$ is similar and thus omitted.

\bigskip
\noindent\textit{Proof of Theorem \ref{consistency}.}
Let $b^{(m)}=\sum_{j=1}^mb_j\phi_j$, then $||b^{(m)}-b||\rightarrow 0$ as $m\rightarrow\infty$.
During the proof for consistency of functional linear models, \cite{cardot99} showed that 
\begin{equation}\label{e1}
||\hat{b}-b^{(m)}||\le C(\frac{1}{\lambda_m^2}+\frac{\sum_{j=1}^m\delta_j^{-1}}{\lambda_m})||\Delta||\cdot||\hat{S}-S||+\frac{2}{\lambda_m}||\hat{\Delta}-\Delta||
\end{equation}
on the event $\{|\hat{\lambda}_m-\lambda_m|\le\lambda_m/2\}$. 

For any $\epsilon>0$, we have 
\begin{eqnarray}\label{e2}
&&P(||\hat{S}-S||>(\frac{2}{\lambda_m^2}+\frac{6\sum_{j=1}^m\delta_j^{-1}}{\lambda_m})^{-1}\epsilon)\nonumber\\
&=&P(k_n^{-1}||\hat{S}-S||>k_n^{-1}(\frac{2}{\lambda_m^2}+\frac{6\sum_{j=1}^m\delta_j^{-1}}{\lambda_m})^{-1}\epsilon)\rightarrow 0
\end{eqnarray}
using Lemma \ref{lem1} since $k_n^{-1}(2/\lambda_m^2+6\sum_{j=1}^m\delta_j^{-1}/\lambda_m)^{-1}\rightarrow\infty$ by assumptions (F) and (G). Similarly we have 
\begin{equation}\label{e3}
P(||\hat{\Delta}-\Delta||>\lambda_m\epsilon)\rightarrow 0.
\end{equation}
Finally, 
\begin{equation}\label{e4}
P(\{|\hat{\lambda}_m-\lambda_m|>\lambda_m/2\})\le P(||\hat{S}-S||>\lambda_m/2)\rightarrow 0
\end{equation}
 by Lemma \ref{lem1} and assumption (F). Equations (\ref{e1})-(\ref{e4}) together imply the consistency result for $\hat{b}$.

For $|\hat{g}(t)-g(t)|$, one only need to note that
\begin{equation}\label{nprate}
|\hat{g}(t)-g(t)|\le |\hat{g}^*(t)-g(t)|+||\sum_iw(t,T_i)X_i||\cdot||\hat{b}-b||,
\end{equation}
where $\hat{g}^*(t)=\sum_iw(t,T_i)(g(T_i)+\epsilon_i)$. The by now standard results in \cite{ferraty04,ferraty06} tell us $|\hat{g}^*(t)-g(t)|=O_p(k_n)$.

\bigskip
\noindent\textit{Proof of Theorem \ref{rate}.}
In the proof, $C$ denotes a generic constant that can assume different values at different places it appears. First we note that directly using equation (\ref{e1}) results in slower rate $||\hat{b}-b||^2=O_p(k_n^2m^{4\alpha+4}+m^{-2\beta+1})$. Instead, we use the decomposition bound
\begin{eqnarray*}
&&||\sum_{j=1}^m\hat{b}_j\hat{\phi}_j-b||^2\\
&\le& C\sum_{j=1}^m(\hat{b}_j-b_j)^2+||\sum_{j=1}^mb_j\hat{\phi}_j-b||^2\\
&\le&C\left(\sum_{j=1}^m(\frac{\langle\hat{\Delta},\hat{\phi}_j\rangle}{\hat{\lambda}_j}-\frac{\langle\Delta,\phi_j\rangle}{\hat{\lambda}_j})^2+\sum_{j=1}^m(\frac{\langle\Delta,\phi_j\rangle}{\hat{\lambda}_j}-\frac{\langle\Delta,\phi_j\rangle}{\lambda_j})^2\right.\\
&&+\left.\int[\sum_{j=1}^mb_j(\hat{\phi}_j-\phi_j)]^2+\sum_{j=m+1}^\infty b_j^2\right)\\
&=:&A_1+A_2+A_3+A_4.
\end{eqnarray*}
On the event $\{|\hat{\lambda}_j-\lambda_j|\le\lambda_j/2,j\le m\}$ which happens with probability converging to $1$, and using the fact $|\hat{\lambda}_j-\lambda_j|\le||\hat{S}-S||$ and $||\hat{\phi}_j-\phi_j||\le2\sqrt{2}||\hat{S}-S||/\delta_j$ \cite{bhatia83,hall07}, where $\delta_j$ is defined in assumption (G), together with Lemma \ref{lem1}, we have
\begin{eqnarray*}
A_1&\le& C\sum_{j=1}^m\lambda_j^{-2}\left[\langle\hat{\Delta}-\Delta,\hat{\phi}_j\rangle^2+\langle\Delta,\hat{\phi}_j-\phi_j\rangle^2\right]\\
&\le& \sum_{j=1}^mj^{2\alpha}j^{2\alpha+2}k_n^2=O_p(k_n^2m^{4\alpha+3}),\\
A_2&\le&C\sum_{j=1}^m\frac{(\hat{\lambda}_j-\lambda_j)^2}{\lambda_j^4}\langle\Delta,\phi_j\rangle^2=C\sum_{j=1}^m\frac{(\hat{\lambda}_j-\lambda_j)^2}{\lambda_j^4}\lambda_j^2b_j^2\\
&=&C\sum_{j=1}^m\frac{b_j^2}{\lambda_j^2}(\hat{\lambda}_j-\lambda_j)^2\le C\sum_{j=1}^mj^{2\alpha-2\beta}k_n^2=O_p(k_n^2m^{2\alpha+1}),\\
A_3&\le& Cm\sum_{j=1}^mb_j^2||\hat{\phi}_j-\phi_j||^2\le Cm\sum_{j=1}^m\frac{b_j^2}{\delta_j^2}k_n^2\\
&\le& Cm\sum_{j=1}^mj^{2\alpha-2\beta+2}k_n^2=O_p(k_n^2m^{2\alpha+2}),\\
A_4&\le& C\sum_{j=m+1}^\infty j^{-2\beta}=O_p(m^{-2\beta+1}).
\end{eqnarray*}
The conclusion $||\hat{b}-b||^2=O_p(k_n^2m^{4\alpha+3}+m^{-2\beta+1})$ now directly follows from the above bounds for $A_i, i=1,2,3,4.$ 

Finally, the convergence rate for the nonparametric component follows directly from (\ref{nprate}).

\end{document}